\documentclass[a4paper,12pt]{article}
\newcommand{\doublespacing}{\let\CS=\@currsize\renewcommand{\baselinestretch}
{1.5}\tiny\CS}
\newcommand{\lyxaddress}[1]{
  \par {\raggedright #1
  \vspace{1.4em}
  \par}
}

\begin{document}
\title{No-Flipping as a consequence of No-Signalling and
Non-increase of Entanglement under LOCC}

\author{Indrani Chattopadhyay \protect\( ^{1}\protect \)\
\thanks{ichattopadhyay@yahoo.co.in} ,
Sujit K. Choudhary\protect\( ^{2}\protect \)\
\thanks{sujit$\_$r@isical.ac.in} , Guruprasad Kar\protect\( ^{2}\protect \)\
\thanks{gkar@isical.ac.in} , \\ Samir Kunkri\protect\( ^{2}\protect \)\
\thanks{skunkri$\_$r@isical.ac.in} \ and Debasis Sarkar\protect\(
^{1}\protect \)\
\thanks{dsappmath@caluniv.ac.in}}
\maketitle

{ \lyxaddress{\protect\( ^{1}\protect \)Department of Applied
Mathematics, University of Calcutta, 92 A. P. C. Road, Kolkata-
700 009, India}}

{ \lyxaddress{\protect\(^{2}\protect \)Physics and Applied
Mathematics Unit, Indian Statistical Institute, 203 B. T. Road,
Kolkata- 700 108, India}}

\begin{abstract}
Non existence of Universal Flipper for arbitrary quantum states is
a fundamental constraint on the allowed operations performed on
physical systems. The largest set of qubits that can be flipped by
a single machine is a great circle of the Bloch-sphere. In this
paper, we show the impossibility of universal exact-flipping
operation, first by using the fact that no faster than light
communication is possible and then by using the principle of
``non-increase of entanglement under LOCC". Interestingly, in both
the cases, there is no violation of the two principles if and only
if the set of states to be flipped, form a great circle.

PACS number(s): 03.67.Mn, 03.67.Hk

Keywords: Flipping, Entanglement, No-signalling.

\end{abstract}

\vspace{1cm}

The structure of the allowed operations performed on the quantum
systems imposes some restrictions on the systems. Sometimes these
restrictions play a crucial role to understand the basic features
of the system and naturally our task is to find the fundamental
nature of these restrictions in a simple way. It has been shown
that an arbitrary state taken from a set of two known,
non-orthogonal states can not be copied exactly in a deterministic
way {\cite{wootters}}. Similarly, one can not delete copy of an
unknown state by performing some linear, trace preserving joint
operations on two copies of that state {\cite{pati11, zurek}}.
Interestingly, several authors derived these no-cloning and
no-deleting theorems by applying some fundamental principles of
nature like impossibility of signalling {\cite{gisin1, hardy99,
pati00, delet}}, preservation  of entanglement for closed systems
under local operations {\cite{horo}} or rather increase of
entanglement by LOCC {\cite{rev}}.

Another interesting feature of quantum system, is the
non-existence of universal flipping machine {\cite{unot}} for
arbitrary input qubit states, {\emph{i.e.,}} there exists no
universal flipper which can operate on any unknown qubit state
$|\psi\rangle$ resulting the orthogonal state $|\psi^\perp\rangle$
{\cite{maspop, massar, gisin, Enk}}. This no-flipping theorem has
a stark dissimilarity with others, as unlike no-cloning and
no-deleting, two non-orthogonal states can always be flipped.
Actually  the largest set of states (of qubit system) which can be
flipped exactly, by a single unitary operator is the set of states
lying on a great circle of the Bloch sphere {\cite{ghosh, patire,
pati}}.

In this paper our aim is to establish the no-flipping theorem by
applying the following established principles of nature:

\emph{1. Impossibility of superluminal signalling - It is
impossible to communicate any message between some spatially
separated parties with a speed greater than the speed of light.}

\emph{2. The thermodynamical law of Entanglement - Amount of
Entanglement shared between some spatially separated parties can
not be increased by LOCC, \emph{i.e.,} by performing local
operations on the subsystems and classical communications between
them.}

In other words our aim is to show, if exact flipping of even the
minimal number (\emph{i.e.,} three) of states, not taken from one
great circle, is possible, then one can send instantaneous signal
as well as increase  entanglement between two distant parties by
local operations.

For this purpose we consider three arbitrary states not lying in
one great circle in their simplest form as;
\begin{quote}
\begin{equation}
\begin{array}{lcl}
|0\rangle, \\
|\psi\rangle= a|0\rangle + b|1\rangle, \\ |\phi\rangle= c|0\rangle
+ d~e^{i\theta}|1\rangle,\end{array}
\end{equation}
\end{quote}
where a, b, c, d are real numbers satisfying the relation $ ~ a^2
+ b^2 ~=~ 1 ~=~ c^2 + d^2~$ and $0 < \theta < \pi, a>0, c>0$, and
the states $|0\rangle$ , $|1\rangle$ are orthogonal to each other.

We assume that a machine exists which can flip at least these
three states exactly. The most general flipping operation for
these three states can be described as

\begin{equation}
\begin{array}{lcl}
& & |0\rangle|M\rangle \longrightarrow |1\rangle |M_{0}\rangle \\
& & |\psi\rangle|M\rangle \longrightarrow
e^{i\mu}|\overline{\psi}\rangle |M_{\psi}\rangle \\

& &  |\phi\rangle|M\rangle \longrightarrow
e^{i\nu}|\overline{\phi}\rangle |M_{\phi}\rangle
\end{array}
\end{equation}

where $\mu$ and $\nu$ are some arbitrary phases and $|M\rangle$ is
the initial machine state. The flipped states are orthogonal to
the original states, \emph{i.e.,}

\begin{equation}
\begin{array}{lcl}
& & \langle 0 | 1 \rangle = \langle \psi | \overline{\psi} \rangle
= \langle \phi | \overline{\phi} \rangle = 0,
\end{array}
\end{equation}
where $| \overline{\psi} \rangle = b| 0 \rangle - a | 1 \rangle$
and $ | \overline{\phi} \rangle = d~e^{-i\theta}|0\rangle -
c|1\rangle$ in their usual notations.

 The above operation is not assumed to be unitary and the
operation acts linearly on one side of an entangled state only
when the density matrix has a mixture representation of the states
in equation $(2)$.

First we show that the exact flipping machine which we have
considered, implies signalling. Consider two spatially separated
parties, say, Alice and Bob who initially share an entangled state
of the form,
\begin{equation}
\begin{array}{lcl}
|\Psi\rangle^{i}_{AB}&=&~\frac{1}{\sqrt{3}}~ (|0\rangle_A
|0\rangle_B + |1\rangle_A |\psi\rangle_B + |2\rangle_A |
\phi\rangle_B)\otimes~|M\rangle_B
\end{array}
\end{equation}

where Alice holds a system associated with three dimensional
Hilbert space, having a basis, $\{~|0\rangle,~|1\rangle,~|2\rangle
\}$(say) and Bob's system consists of a qubit (entangled with
Alice's system) and  a flipping machine defined as in equation
(2). Here one should note that the joint system of Alice and Bob
has been chosen in such a manner that the marginal density matrix
of Bob's side admits a representation in terms of the three states
$|0\rangle,~|\psi\rangle,~|\phi\rangle$, on which the flipping
machine has been defined.

The reduced density matrix of Alice's side is
\begin{equation}
\begin{array}{lcl}
\rho_A^i &=& \frac{1}{3}~
\{~P[|0\rangle]+P[|1\rangle]+P[|2\rangle]+
a(|0\rangle\langle1|+|1\rangle\langle0|)\\ &
&+c(|0\rangle\langle2| +|2\rangle\langle0|)+
\langle\phi|\psi\rangle |1\rangle\langle2| +
\langle\psi|\phi\rangle|2\rangle\langle1|~\}
\end{array}
\end{equation}

Now assume that Bob applies the flipping machine on his qubit.
After the flipping operation the shared state between Alice and
Bob takes the following form,
\begin{equation}
\begin{array}{lcl}|\Psi\rangle^{f}_{AB}=~\frac{1}{\sqrt{3}}~
\{|0\rangle_A |1~M_0\rangle_B +
e^{i\mu}|1\rangle_A |\overline{\psi}~M_{\psi}\rangle_B +
e^{i\nu}|2\rangle_A |\overline{\phi}~M_{\phi}\rangle_B\}
\end{array}
\end{equation}

The final density matrix of Alice's side (expanded in the
computational basis) is
\begin{equation}
\begin{array}{lcl}
\rho^{f}_A &=& ~\frac{1}{3}~ \{P[|0\rangle] +
P[|1\rangle]+P[|2\rangle]- a(e^{-i\mu}\langle M_\psi | M_0\rangle
|0 \rangle\langle 1|\\& &+ e^{i\mu}\langle M_0 | M_\psi\rangle
|1\rangle\langle 0|) -c(e^{-i\nu}\langle M_\phi | M_0\rangle
|0\rangle\langle 2|+ e^{i\nu} \langle M_0 | M_\phi\rangle
|2\rangle\langle 0|)\\& & + \langle\psi|\phi\rangle
e^{i(\mu-\nu)}\langle M_\phi | M_\psi \rangle
 |1\rangle \langle 2|+
\langle\phi|\psi\rangle e^{i(\nu-\mu)}\langle M_\psi | M_\phi
\rangle |2\rangle\langle1|\}
\end{array}
\end{equation}

As this is a trace preserving local operation performed entirely
on Bob's side and there is also no classical communication between
them, so to prevent any violation of the principle of
no-signalling, the reduced state on Alice's side must remain
unchanged. Equating the  reduced density matrices on Alice's side
before and after the flipping operation on Bob's side we get the
following relations

\begin{equation}
a~=~ -a e^{i\mu}\langle M_0 | M_\psi\rangle~=~ -ae^{-i\mu}~\langle
M_\psi | M_0\rangle
\end{equation}
\begin{equation}
c~=~ -c e^{i\nu}\langle~ M_0 | M_\phi\rangle~=~
-ce^{-i\nu}\langle~ M_\phi | M_0\rangle
\end{equation}
\begin{equation}
\langle\phi|\psi\rangle~=~e^{i(\mu-\nu)}~\langle\psi|\phi\rangle
~\langle M_\phi | M_\psi \rangle
\end{equation}
\begin{equation}
\langle\psi|\phi\rangle~=~e^{i(\nu-\mu)}~\langle\phi|\psi\rangle
~\langle M_\psi | M_\phi \rangle
\end{equation}

The above relations imply,
$$ \mbox{either}~b=0~\mbox{or}~d=0~\mbox{or}~\sin{\theta}=0,$$
forcing the states to lie on a great circle. So, we observe that
as long as the three states do not lie on a great circle, we have,
$\rho^i_A \neq \rho^f_A$, and interestingly whenever the equality
holds, i.e., $\rho^i_A~=~\rho^f_A$, the three states actually lie
on a great circle. This clearly shows that exact flipping of any
three states not lying on a great circle is an impossibility.

Now, we are going to show the impossibility of universal flipping
using the principle of non-increase of entanglement by LOCC. Here
we must be careful about the fact that in the earlier set-up
(equation $(4)$), eigen values of $\rho^i_A$ and $\rho^f_A$ are
equal, implying no change in entanglement even when states are not
on the great circle. For this we choose another state,
$|\Psi^i_{AB}\rangle$ of five qubits, shared between Alice and
Bob, situated at distant locations, where the first qubit is with
Alice, and remaining four($B_1, B_2, B_3, B_4$) are with Bob.
\begin{equation}
\begin{array}{lcl}
|\Psi^{i}_{AB}\rangle &=& \frac{1}{\sqrt{8}}~\{
(|000\rangle+|111\rangle)_{AB_1B_2}
\otimes{|10\rangle}_{B_3B_4}\\& &
-(|010\rangle+|100\rangle+|101\rangle)_{AB_1B_2}\otimes{|\overline{\psi}
\psi\rangle}_{B_3B_4}\\& & -
(|011\rangle+|110\rangle+|001\rangle)_{AB_1B_2}\otimes
{|\overline{\phi}\phi \rangle}_{B_3B_4}\} \otimes
{|M\rangle}_{B_M}
\end{array}
\end{equation}

Let us assume that Bob  has a flipping machine (defined earlier)
with him which he applies on his last qubit($B_4$). After the
flipping operation the joint state between them takes the form,
\begin{equation}
\begin{array}{lcl}
|\Psi^{f}_{AB}\rangle &=& ~
\frac{1}{\sqrt{8}}~\{(|000\rangle+|111\rangle)_{AB_1B_2}
\otimes{|11M_0\rangle}_{B_3B_4B_M}\\& &- e^{i\mu}
(|010\rangle+|100\rangle+|101\rangle)_{AB_1B_2}\otimes
{|\overline{\psi}~ \overline{\psi} M_{\psi}
\rangle}_{B_3B_4B_M}\\& &-
e^{i\nu}(|011\rangle+|110\rangle+|001\rangle)_{AB_1B_2}\otimes{|\overline{\phi}~
\overline{\phi} M_{\phi} \rangle}_{B_3B_4B_M}\}.
\end{array}
\end{equation}

In order to compare the amount of entanglement present in
$|\Psi^{i}_{AB}\rangle$ with that present in
$|\Psi^{f}_{AB}\rangle$, we compare the eigenvalues of the
marginal density matrices on Alice's side before and after the
flipping operation.

The initial state of Alice's subsystem is;
\begin{equation}
\begin{array}{lcl}
\rho^{i}_{A}&=&
\frac{1}{8}\{4(P[|0\rangle]+P[|1\rangle])+(a^2+c^2+2|\langle
\psi|\phi\rangle|^2)(|0 \rangle \langle 1|+|1\rangle \langle
0|)\}.
\end{array}
\end{equation}

The final state of Alice's subsystem is;
\begin{equation}
\begin{array}{lcl}
\rho^{f}_{A}&=&~\frac{1}{8}\{4(P[|0\rangle]+P[|1\rangle])+(2Z-a^2X^*-c^2Y)|0\rangle
\langle 1|\\& &~ +(2Z-a^2X-c^2Y^*)|1\rangle \langle 0|\}
\end{array}
\end{equation}

where,  $Z=Re[e^{i(\mu-\nu)}(\langle \psi| \phi\rangle)^2 \langle
M_\phi| M_\psi\rangle],~X=e^{i \mu}~\langle M_0|M_\psi\rangle $
and\\ $Y= e^{i \nu}~\langle M_0|M_\phi\rangle .$

After constructing the eigenvalue equations for both
$\rho^{i}_{A}$ and $\rho^{f}_{A}$ we find that the largest
eigenvalues are respectively

$$\lambda^i = \frac{1}{2}+\frac{1}{8}(~2|\langle
\psi|\phi\rangle|^2+a^2+c^2)$$ and $$\lambda^f
=\frac{1}{2}+\frac{1}{8}|~2Z-a^2X^*-c^2Y|$$

It can easily be checked that $\lambda^f \leq \lambda^i$( see
appendix). This implies, $E(|\Psi^{f}_{A:B}\rangle) \geq
E(|\Psi^{i}_{A:B}\rangle)$ where $E$ stands for amount of
entanglement. So this establishes the impossibility of universal
flipping as here the only operation that has been allowed is
local. A close look will reveal that  the greatest eigenvalues are
equal (i.e., no increase of entanglement), only when, the three
states on which we have defined our flipping machine, lie on one
great circle.

The above construction seems to be a complicated one due to our
linearity assumption. We have considered our flipping
operation acts linearly on one side of an entangled state only
when the reduced density matrix has a mixture representation of the states
in equation $(2)$. However a more simpler proof is possible consisting
only a three qubit system if we allow further that the operation
is linear on superposition level.

Now consider a three qubit state shared between Alice and Bob
where Bob has a two qubit system ($B_1$ and $B_2$) as follows,
\begin{equation}
\begin{array}{lcl}
|\Phi^i \rangle _{AB} = \frac{1}{\sqrt{b^2+d^2}}\{~|0\rangle _A
{\frac{|0\rangle _{B_1} |\psi \rangle _{B_2}-|\psi\rangle _{B_1}
|0\rangle_ {B_2}}{\sqrt{2}}} ~+~ |1\rangle _A  \frac{|0\rangle
_{B_1}|\phi \rangle_ {B_2}-|\phi \rangle _{B_1} |0\rangle_{B_2}
}{\sqrt{2}}\}\\
=\frac{1}{\sqrt{b^2+d^2}}\{(b|0\rangle _A + d e^{i\theta}
|1\rangle _A) \frac{|0\rangle _{B_1}|1 \rangle_ {B_2}-|1 \rangle
_{B_1} |0\rangle_{B_2} }{\sqrt{2}}\}
\end{array}
\end{equation}
Clearly $|\Phi^i \rangle_{AB}$ is a separable (pure product) state
in $A:B$ cut. Assume Bob has a flipping machine with him which he
applies on his last qubit, i.e., on $B_2$. Then the joint state
between them takes the form,
\begin{equation}
\begin{array}{lcl}
|\Phi^f \rangle _{AB} =
\frac{1}{\sqrt{N}}\{e^{i\mu}|00\rangle|\overline{\psi}\rangle
|M_{\psi}\rangle+ e^{i\nu}|10\rangle|\overline{\phi}\rangle
|M_{\phi}\rangle-(|0\rangle |\psi \rangle + |1\rangle |\phi
\rangle)|1 \rangle |M_0 \rangle \}
\end{array}
\end{equation}
where $N=2+ a^2 Re\{e^{-i\mu} \langle M_\psi \mid M_0 \rangle \}+
c^2 Re\{e^{-i\nu} \langle M_\phi \mid M_0 \rangle\}$. It is easy
to check that the state in general is an entangled state in $A:B$
cut, thus in this setting we also observe an increase of entanglement
by LOCC.

In the conclusive remarks, we want to mention that, as a
constraint on quantum mechanical system, the `No-Flipping' theorem
is weaker than the `No-Cloning' and `No-Deleting' theorem, because
exact flipping is possible for all states lying in one great
circle of the Bloch sphere. Hence, showing the impossibility of
universal flipping from the principle like No-signalling and Non
increase of entanglement by local operations, is an interesting
problem to deal with. We find the nonphysical nature of flipping
operation on whole Bloch sphere in two different ways. In this
context, recently no-flipping has been established by using the
existence of incomparable states where one qutrit and two qubits
are required {\cite{incompare}}.

\begin{center}
\textit{Appendix}
\end{center}
To check that, $\lambda^f \leq \lambda^i$ for the second set-up,
we have,\\ $\lambda^{i} \geq \lambda^{f}$
$\Leftrightarrow~2|\langle \psi|\phi\rangle|^2+a^2+c^2\geq |~2Z-a^2X^*-c^2Y|$\\
$\Leftrightarrow ~a^4(1-|X|^2)+c^4(1-|Y|^2)+ 2a^2
c^2[1-Re(XY)]+4a^2[|\langle \psi|\phi\rangle|^2+Z
Re(X)]+4c^2[|\langle \psi|\phi\rangle|^2+Z Re(Y)]
+4[|\langle \psi|\phi\rangle|^4 -Z^2]\geq 0$\\
As, all terms on the L.H.S. of the last inequality are
non-negative. Therefore, the above inequality for eigenvalues is
satisfied.

The equality holds only when, $~|X|=|Y|=1, ~Re(XY)=1, ~|\langle
\psi|\phi\rangle|^2+Z Re(X)=0, |\langle \psi|\phi\rangle|^2+Z
Re(Y)=0,~
Z^2= |\langle \psi|\phi\rangle|^4.$\\
Now, $Z=Re[e^{i(\mu-\nu)}(\langle \psi| \phi\rangle)^2 \langle
M_\phi| M_\psi\rangle],~X=e^{i \mu}~\langle M_0|M_\psi\rangle $
and \\ $Y= e^{i \nu}~\langle M_0|M_\phi\rangle,$ which implies
machine states will differ by only some phases and the states
$|0\rangle,|\psi\rangle,|\phi\rangle$ will lie on a great circle.

Hence we see that in this case, where $\rho^{i}_{A}=\rho^{f}_{A}$,
the three states of equation (1) on which we have defined our
flipping machine, actually lies on a great circle of the Bloch
sphere.\\

{\bf Acknowledgement.} We thank Prof. N. Gisin for his valuable
comments and suggestions.  Also we acknowledge the referee for
his/her valuable comments and suggestions. I.C. and S.K. also
acknowledges CSIR, India for providing fellowship during this work.\\


\begin{thebibliography}{99}
\bibitem{wootters} W. K. Wootters and W. H. Zurek, {\it Nature} {\bf 299}, 802 (1982).
\bibitem{pati11} A. K. Pati and S. L. Braunstein, {\it Nature} {\bf 404}, 164
(2000).
\bibitem{zurek} W. H. Zurek, {\it Nature} {\bf 404}, 40 (2000).
\bibitem{gisin1} N. Gisin, {\it Phys. Lett. A} {\bf 242}, 1-3 (1998).
\bibitem{hardy99} L. Hardy and D. D. Song, {\it Phys. Lett. A} {\bf 259}, 331(1999).
\bibitem{pati00} A. K. Pati, {\it Phys. Lett. A} {\bf 270}, 103
(2000).
\bibitem{delet} A. K. Pati and S. L. Braunstein, {\it Phys. Lett. A} {\bf 315}, 208-212 (2003).
\bibitem{horo} M. Horodecki, R. Horodecki, A. Sen(De) and U. Sen,
arXiv:quant-ph/0306044.
\bibitem{rev} I. Chattopadhyay, S. K. Choudhary, G. Kar, S. Kunkri and D. Sarkar, in preparation.
\bibitem{unot} V. Buzek, M. Hillery and R. F. Werner, {\it Phys. Rev. A} {\bf 60},
R2626-R2629 (1999).
\bibitem{maspop} S. Massar and S. Popescu, {\it Phys. Rev. Lett.} {\bf 74}, 1259 (1995).
\bibitem{massar} S. Massar, {\it Phys. Rev. A} {\bf
62}, 040101(R) (2000).
\bibitem{gisin} N. Gisin and S. Popescu, {\it Phys. Rev. Lett.} {\bf 83}, 432-435
(1999).
\bibitem{Enk} S. J. van Enk, {\it Phys. Rev. Lett.} {\bf 95}, 010502 (2005).
\bibitem{ghosh} S. Ghosh, A. Roy and U. Sen {\it Phys. Rev. A} {\bf 63}, 014301 (2000).
\bibitem{patire} A. K. Pati, {\it Phys. Rev. A} {\bf 63}, 014302 (2000).
\bibitem{pati} A. K. Pati, {\it Phys. Rev. A} {\bf 66}, 062319 (2002).
\bibitem{incompare} I. Chattopadhyay and D. Sarkar, arXiv:quant-ph/0511040.


\end{thebibliography}
\end{document}